# Designing 3D topological insulators by 2D-Xene ($X$ = Ge, Sn) sheet functionalization in the GaGeTe-type structures

F. Pielnhofer,*[a] T. V. Menshchikova,*[b] I. P. Rusinov,[b,c] A. Zeugner,[d] I. Yu. Sklyadneva[b,e,f,g], R. Heid[f], K.-P. Bohnen[f], P. Golub,[d] A. I. Baranov,[d,h] E. V. Chulkov,[b,c,g,i,j] A. Pfitzner,[a] M. Ruck[d,h] and A. Isaeva*[d]

a. University of Regensburg, Institute of Inorganic Chemistry, Universitätsstr. 31, 93053 Regensburg, Germany. Email: florian.pielnhofer@ur.de

b. Tomsk State University, pr. Lenina, 36, 634050 Tomsk, Russia. Email: menshikova_t@mail.ru

c. St. Petersburg State University, Universitetskaya nab., 7/9, 199034 St. Petersburg, Russia.

d. Technische Universität Dresden, Department of Chemistry and Food Chemistry, Helmholtzstraße 10, 01069 Dresden, Germany. Email: anna.isaeva@tu-dresden.de

e. Institute of Strength Physics and Materials Science, pr. Academicheskii 2/1, 634021, Tomsk, Russian Federation

f. Karlsruher Institut für Technologie, Institut für Festkörperphysik, D-76021 Karlsruhe, Germany

g. Donostia International Physics Center (DIPC), Paseo de Manuel Lardizabal, 4, 20018 San Sebastián/Donostia, Basque Country, Spain.

h. Max Planck Institute for Chemical Physics of Solids, Nöthnitzer Str. 40, 01187 Dresden, Germany.

i. Departamento de Física de Materiales, Facultad de Ciencias Químicas, UPV/EHU, 20080 San Sebastián/Donostia, Basque Country, Spain.

j. Centro de Física de Materiales CFM–MPC, Centro Mixto CSIC–UPV/EHU, 20080 San Sebastián/Donostia, Basque Country, Spain.

ABSTRACT State-of-the-art theoretical studies anticipate a 2D Dirac system in the "heavy" analogues of graphene, free-standing buckled honeycomb-like *X*enes ($X$ = Si, Ge, Sn, Pb, etc.). Herewith a structurally and electronically resembling 2D sheet, which can be regarded as *X*ene functionalized by covalent interactions within a 3D periodic structure, is predicted to constitute a 3D strong topological insulator with $Z_2 = 1;(111)$ (primitive cell, rhombohedral setting) in the structural family of layered *AX*Te ($A$ = Ga, In; $X$ = Ge, Sn) bulk materials. The host structure GaGeTe is a long-known bulk semiconductor; the "heavy", isostructural analogues InSnTe and GaSnTe are predicted to be dynamically stable. Spin-orbit interaction in InSnTe opens a small topological band gap with inverted gap edges that are mainly composed of the In-$5s$ and Te-$5p$ states. Our simulations classify GaSnTe as a semimetal with topological properties, whereas the verdict for GaGeTe is not conclusive and urges further experimental verification. *AX*Te family structures can be regarded as stacks of 2D layered cut-outs from a zincblende-type lattice and are composed by elements that are broadly used in modern semiconductor devices; hence they represent an accessible, attractive alternative for applications in spintronics. The layered nature of *AX*Te should facilitate exfoliation of its hextuple layers and manufacture of heterostuctures.

**INTRODUCTION**

Surface properties originating from the global and crystal-lattice symmetries have attracted a great deal of attention in the past decade.[1] This interest may be fuelled in the foreseeable future by the Nobel Prize in Physics awarded in 2016 for the discovery of topological phases of matter and topological transitions. Materials hosting 2D and 3D Dirac fermions are believed to foster new types of devices and to complement or even excel classic semiconductor transistors. Over just a few years various flavors of topological materials, e. g. topological insulators,[2] topological crystalline insulators and superconductors,[3] non-symmorphic crystalline insulators,[4] Weyl semimetals,[5,6] etc. have been discovered. Herewith we suggest a new platform for 3D strong topological insulators: the GaGeTe-type layered bulk materials that are structurally related to both basic zincblende-type semiconductors and 2D-*X*ene materials[7].

The progenitor GaGeTe has been synthesized as bulk crystals.[8,9] It has a layered crystal structure stacked from six-atom-thick $^2_\infty$[Te–Ga–Ge–Ge–Ga–Te] building blocks (denoted henceforward as a hextuple layer of GaGeTe) separated by van der Waals gaps. Each hextuple layer can be considered as a buckled two-atom-thick germanium sheet in the armchair configuration wrapped in a four-atom-thick structural fragment of the *β*-GaSe-type structure[10].

Whereas further relevant structural peculiarities of GaGeTe are detailed in the next section, the immediate discussion focuses on the corrugated germanium fragment. It bears striking structural similarity to germanene[11] and other 2D monolayers of group IVA atoms (graphene,[12] silicene,[13,14] stanene[15]) that are under intense spotlight nowadays due to the high mobility of charge carriers and are envisioned as components of future transistors. These artificial 2D materials coined *X*enes (*X* = IVA element), which accommodate *X* atoms in the buckled honeycomb arrangement, are predicted to exhibit the quantum spin Hall effect (QSHE), possibly even persisting up to room temperature.[7]

Furthermore, some proposals advocate that topological states emerge in the covalently functionalized *X*ane derivatives. For instance, a 2D topological insulator is expected in halogen-functionalized germanane Ge*X* (*X* = H, F, Cl, Br), methyl-substituted GeCH$_3$[16–18] and ethynyl-derivative of germanene GeC$_2$*X* (*X* = H, halogen)[19] under sizeable tensile strain. Ethynyl- or methyl-functionalized stanene[20,21] and halide-functionalized plumbene[22] exemplify the case of heavier elements. On the other hand, ionically functionalized *X*ene-like structural fragments in Zintl compounds *MX*$_2$ (*M* = Ca, Sr, Ba; *X* = Si, Ge, Sn) may account for an entire family of topological materials ranging from topological nodal-line semimetals to presumably Dirac semimetals and even a strong topological insulator with $Z_2 = 1;(001)$ in BaSn$_2$, as has been very recently found by first-principles calculations.[23–25]

Experimental confirmation of these perspectives has been so far strongly challenged.[7] An impressive achievement is the recently reported synthesis of germanane GeH, a hydrogen-saturated analogue of graphane[26,27], that has been obtained via hydrolysis of bulk *β*-CaGe$_2$ precursor.[28,29] GeH is a trivial wide-gap semiconductor with the band gap of 1.56 eV,[16] and its electronic structure can be flexibly varied by chemical pressure so that the band gap size changes by ca. 15 %.[30]

Herewith we demonstrate by means of a first-principles study that covalent functionalization of an *X*ene-like structural fragment may implicate topological order in the bulk GaGeTe-type structure. Up until now scarce characterization of the physical properties[31–33] and absence of any band-structure calculations have kept GaGeTe away from the mainstream research. We aim to fill in this gap and to entice further experimental verification of the predicted properties.

The present contribution focuses on the electronic structures of bulk GaGeTe and its hypothetical, isostructural analogues, GaSnTe and InSnTe, with stronger spin-orbit coupling. Whereas the latter compounds are predicted to be topological materials on all levels of theory applied (DFT, screened hybrid functional, GW correction), the case of the forerunner remains inconclusive. Being a narrow-gap TI within DFT, GaGeTe is rendered a trivial semiconductor with a much larger band gap by HSE06 and the GW-approach. Lately, theory has helped to identify many TI candidate materials with the aid of the $Z_2$ classification,[34–37] and ensuing experiments confirmed or disproved these predictions for a considerable number of "contenders"[2]. In the course of that pursuit, the problem of false-positive TI prospects churned out by DFT calculations was identified and the rather resilient GW-method was proposed to ameliorate it[38,39]. Noteworthy, the hybrid HSE functional, which is traditionally regarded as superior to the standard DFT ones, was also found to yield false-negative results in the search of new TIs, as opposed to DFT and GW[38]. Thus, the contradicting theoretical predictions for GaGeTe urge experimental efforts such as transport measurements and spectroscopy studies for the ultimate clarification.

**METHODS**

**Electronic structures**

Electronic structure calculations were carried out within the framework of density functional theory (DFT). Various program packages were used complementary in order to verify the obtained electronic properties.

Structural optimizations and calculations of the *AX*Te (*A* = Ga, In; *X* = Ge, Sn) band structures were performed with the projector augmented-wave (PAW) method[40] as implemented in the VASP (Vienna Ab Initio Simulation Package) code[41–43]. With VASP the exchange-correlation energy was treated using the generalized gradient approximation (GGA) with the Perdew–Burke–Ernzerhof (PBE)[44] parametrization. Scalar-relativistic corrections were included into the Hamiltonian and the spin-orbit coupling (SOC) was taken into account by the second variation method[45]. A $k$-point mesh of $7 \times 7 \times 7$ was used after the preliminary tests showed that an increased mesh did not affect the obtained spectra. Bulk relaxation of *AX*Te was carried out by the DFT+D3 method that correctly describes the van der Waals interactions[46,47]. Furthermore, topological character of the *AX*Te electronic structures has been tested by the calculations with the exact exchange functional HSE06[48,49] which includes a Hartree–Fock term in the exchange part. This functional is known to represent band structures of semiconductors with higher accuracy with respect to DFT[50].

$Z_2$ invariants were computed via the parities of the wave functions according to the Fu and Kane formalism[51] and by the method implemented in Z2Pack[52,53]. The results obtained by both approaches are in full agreement.

GW calculations were performed using VASP[41–43] and WANNIER90[54,55] codes. On the first stage, DFT calculations employing the PBE functional were performed without including the spin-orbit coupling. For the calculation of the dielectric function 300 bands were chosen that correspond to the energy window up to 100 eV above the Fermi level. The $k$-point mesh was chosen to be $7 \times 7 \times 7$. The SOC was taken into account using an a posteriori treatment method[39] on the basis of the Wannier interpolations technique.

Furthermore, full structural optimizations were carried out with the linear combination of atomic orbitals (LCAO) method as implemented in CRYSTAL14[56] for GaGeTe as well as for the hypothetical model compounds GaSnTe and InSnTe. Apart from the PBE parametrization, plus Grimme's D2 dispersion correction[57] the local density approximation (LDA) in the Vosko-Wilk-Nusair (VWN)[58] parametrization was applied. The total energy was converged on a $k$-mesh with $10 \times 10 \times 10$ $k$-points. Besides adjusted all-electron basis sets for Ga (86-4111d41G)[59], Ge (97-631d61G)[60], In (97-63111d631G)[61–63] and Sn (97-63111d631G)[64] an all-electron basis set for Te[65] or a pseudo-potential basis (m-pVDZ-PP) for a scalar-relativistic description of Te[66] were used.

Electronic structures of *AX*Te were additionally assessed by the full potential local orbital method (FPLO)[67] as implemented in the FPLO-program (version 14.00-45). The PBE and the LDA with the Perdew–Wang (PW91)[68] parametrizations were applied. For GaGeTe, the experimental structure, the optimized geometries from the CRYSTAL calculations and from the FPLO-LDA approach were considered as an input for band structure calculations. The hypothetic GaSnTe and InSnTe structure models were taken solely from the CRYSTAL calculations. A full-relativistic Hamiltonian (Dirac–Coulomb) was applied in the FPLO calculations and the total energy was converged on a $k$-mesh with $12 \times 12 \times 12$ $k$-points.

Full-potential (L)APW+lo+LO LDA[68] DFT calculations were performed with the ELK code[69]. A scalar-relativistic Hamiltonian after *Koelling* and *Harmon* was used[45]. Spin-orbit coupling was taken into account perturbatively at the second variational step and included only the spherical part of the Kohn–Sham potential inside muffin-tin spheres as implemented in the ELK code. A $k$-mesh of 11 $k$-points inside the irreducible part of the Brillouin zone for the primitive lattice was used. The RGk$_{max}$ parameter and the angular momentum cut-off used for the wave function expansion inside the muffin-tin spheres were chosen equal to 8. Further computational details can be found in the Supplementary Information (Table S6 in ESI).

Dielectric function and corresponding optical coefficients of GaGeTe were also calculated with the ELK code[69] on the $7 \times 7 \times 7$ $k$-point grid. Preliminary tests showed that increased $k$-point mesh and switched-on spin-orbit coupling did not bring in any qualitative changes into the computed dielectric function. The ELK results appeared to be in full accordance with the above mentioned VASP results.

Projector-augmented-wave GGA[44] calculations were performed with the ABINIT code[70]. Modified ABINIT datasets[71] were used and the planewave cut-off energy was equal to 20 a. u. Further information on computational parameters can be found in Table S6 in ESI.

In terms of the electronic structures, the results of Elk and ABINIT were completely in accordance with those obtained by the above-mentioned VASP and FPLO. Hence the results of the former were used further as an input for the analysis of the chemical bonding.

For the calculation of the phonon-dispersion spectra, the electronic structure calculations of GaSnTe and InSnTe were performed in the mixed-basis pseudopotential approach[72] with the exchange and correlation energy functional evaluated within the generalized gradient approximation[44]. Spin-orbit coupling was incorporated into the pseudopotential scheme via Kleinman's formulation and treated fully self-consistently[73]. Elastic moduli were calculated from the obtained phonon spectra. Phonon dispersions were calculated using the linear response technique[74] in combination with the mixed-basis pseudopotential method[75].

**Evaluation of chemical bonding**

Evaluation of quantum theory of atoms in molecules (QTAIM) basins was performed[76] from the electron density computed for $AX$Te ($A$ = Ga, In; $X$ = Ge, Sn) on a discrete grid with a ~0.05 a. u. step with the program DGrid[77]. The same code was used to compute the delocalization indices[78–81] between the QTAIM basins from the (L)APW and PAW results[82,83]. Delocalization indices characterize the degree of electron pair exchange between the basins (two atoms) and can be interpreted as the covalent-bond order[81]. For spinor wavefunctions employed in the calculations including spin-orbit coupling, the delocalization indices were computed according to[84].

Additionally, the QTAIM basins were computed for an optimized bulk structure of GaGeTe from the electron densities calculated from all electron basis sets with the CRYSTAL code and analyzed with TOPOND[85]. The results are fully consistent with those obtained by the above mentioned method.

ELI-D (electron localizability indicator) is a real-space bonding indicator[86,87] that partitions the crystal-lattice space into non-overlapping regions (basins) designating atomic cores, penultimate valence shells, electron lone pairs and regions of pairwise or multicentre bonds. Moreover, integration of the electron density within these basins (similar to the QTAIM concept[88]) allows to quantify the electron count for each bond, while the polarity index ($p$)[89] determines the bond polarity through the ratios between the electronic contributions of all bonding constituents.

**RESULTS AND DISCUSSION**

**Crystal structures of bulk $AX$Te ($A$ = Ga, In; $X$ = Ge, Sn)**

The periodic layered structure of GaGeTe[9] can be understood as a stack of 8.17 Å thick, layered packages with a diamond-like atomic arrangement (Fig. 1). Furthermore, these hextuple layers are stacked with antiphase boundaries, making it impossible to derive the entire GaGeTe bulk structure from a zincblende-type 3D lattice with regular voids in the 6$c$ Wyckoff site. Instead GaGeTe adopts an *ABC* stacking sequence of the hextuple layers along the *c* axis resulting in a trigonal unit cell (sp. gr. $R\bar{3}m$, no. 166), so that the otherwise tetrahedral coordination polyhedron of each Te atom remains incomplete due to the missing vertex (Fig. 1). Like quintuple layers in $Bi_2Te_3$, hextuple layers in GaGeTe are separated by van der Waals gaps of about 3.41 Å (defined as a normal between the adjacent Te atomic planes). The shortest inter-layer Ga···Te distances for the atoms in the eclipsed position are equal to 4.670 Å, while the shortest inter-layer Te···Te distances (van der Waals gap) amount to 4.131 Å. Prominent layered nature of GaGeTe accounts for abundant stacking faults in the crystals of this material.[32]

To the best of our knowledge, isostructural analogues of GaGeTe have not been reported. Since topological order is favoured by stronger spin-orbit coupling, we consider a possibility of isovalent substitutions of germanium and gallium by "heavier" analogues, tin and indium, respectively. Earlier studies of the phase equilibria in the *A*–Sn–Te (*A* = Ga, In) systems revealed only two quasi-binary sections in each system, e. g. $A_2Te_3$–SnTe and *A*Te–SnTe, and one stable ternary compound $Ga_6SnTe_{10}$[90]. Additionally, SnTe-based solid solutions with the rock-salt-type structure are known to incorporate several at. % of indium.

Our structural optimisation of unit cell parameters and atomic positions of the hypothetical GaSnTe and InSnTe under space-group restraint within the DFT-D3 scheme yields plausible interatomic distances and coordination polyhedra (Table 1 and S1 in ESI). The functionalized stanene-like fragment with the interatomic Sn–Sn distances of 2.746 Å (GaSnTe) and 2.784 Å (InSnTe) is compressed in comparison with the optimized free-standing 2D-material[18] (2.88 Å) and resembles more the elemental tin (2.81 Å in α-Sn). On the other hand, it is less stretched out than in $BaSn_2$ (2.919 Å). Degree of buckling in *A*SnTe (Table 1) accords well with the experimental data for stanene on a substrate (ca. 1.2 Å[15]). The $A$–Te distances in $A$SnTe (2.752 Å for $A$ = Ga, 2.901 Å for $A$ = In) are widened in comparison with the typical values for the corresponding $A$Te binaries (cf. Ga–Te 2.64–2.69 Å in GaTe[91], In–Te 2.82 Å in the tetrahedral units in InTe[92]). The most peculiar $A$–Sn bonding contacts in $AX$Te echo the rare examples of polyanionic, helical fragments in NaInSn$_2$,[93] NaGaSn$_2$[94] and NaGaSn$_5$[95]. In these Zintl compounds, indium/gallium and tin atoms occupy mixed atomic sites with (distorted) tetrahedral coordination that reside at the distances of 2.792 Å (NaInSn$_2$), 2.733–2.766 Å (NaGaSn$_2$), and 2.763 Å (NaGaSn$_5$). LiInSn[96] with the zincblende structure accommodates both In and Sn in the mixed anionic site with the interatomic distance of 2.891 Å.

Alongside with the credible crystallochemical features, dynamic stability of the discussed $AX$Te is corroborated by calculations of their phonon-dispersion spectra[97] and elastic moduli. Positive elastic moduli are one of the parameters indicating a dynamically stable compound. This requirement is fulfilled for both GaSnTe ($C_{11}$ = 37.36 GPa, $C_{12}$ = 21.49 GPa, $C_{44}$ = 12.31 GPa, $C'$ = 7.9 GPa) and InSnTe ($C_{11}$ = 36.77 GPa, $C_{12}$ = 25.77 GPa, $C_{44}$ = 9.9 GPa, $C'$ = 5.5 GPa).

To conclude, there is room for justified optimism that the considered "heavy" representatives of the GaGeTe-type could be synthesized by optimized synthetic routes. As the recent example of layered Ge$_4$Se$_3$Te shows, even sliced-and-diced systems like Ge–Te and Ge–Se do have structural novelties to unravel.[98]

**Electronic structures of $AX$Te ($A$ = Ga, In; $X$ = Ge, Sn)**

The bulk band structures of $AX$Te were calculated using a variety of DFT-based codes and exchange-correlation functionals (Table 1). The corresponding geometry optimization data for bulk are summarized in Table S1 in ESI. Note that the following discussion is based on the primitive unit cell (rhombohedral setting) which is outlined in Fig. S1 in ESI. For the structure description (Fig. 1) the conventional unit cell (hexagonal setting) is chosen.

GaGeTe demonstrates a gapless band structure in the scalar-relativistic case (Fig. S2a in ESI). The valence band (VB) maximum predominantly consists of the Te-$5p_z$ states, whereas the minimum of the conduction band (CB) has mainly the Ge-4*s* character. When spin-orbit interaction is taken into account within the PBE functional, the electronic spectrum of GaGeTe gaps out. Four distinct regions with different atomic contributions can be traced in the resultant electronic structure (Fig. 2) and in the projected density of states (Fig. 3a). The top part of the GaGeTe electronic spectrum, the conduction band, is formed by Te *p*-orbitals, Ga and Ge s-orbitals. The next region, the top of the VB, extends from the vicinity of the Fermi level down to ca. −4 eV. This broad continuum is constituted by strongly intermixed *p*-states of the tetrahedrally coordinated Ga, Ge and Te atoms with a predominant contribution from the latter. The third region lies between −4 and −7 eV and is characterized mainly by the *s*-orbitals with a sizeable contribution of the Ga atoms. The last part of the electronic structure at ca. −10 eV is governed by quasi-2D Dirac-cones centered at the W points of the 3D Brillouin zone (Fig. 2a). These states are composed largely of the Ge-4*s* orbitals; hence they can be attributed to Ge bonding within the buckled layer (Fig. 2b). The deeper lying Te-5*s* states represent the non-interacting lone pairs which are typical for layered compounds with van der Waals interactions.

The above described general characteristics of the GaGeTe electronic structure are not affected in the wide energy range by the choice of the exchange-correlation functional or addition of many-body contributions within the GW approximation. The key influence of the chosen functional manifests itself near the Fermi level.

In the framework of the PBE functional, VB and CB are inverted at the T point of the 3D Brillouin zone and a narrow indirect band gap opens (see Fig. 2c). Its size is quite small due to the moderate hybridization of the states (cf. Table 1). Along the other directions of the Brillouin zone there is a sizeable gap of ca. 2 eV. The topological nature of the resultant semiconducting ground state is identified by the calculation of the four topological $Z_2$ invariants $\nu_0;(\nu_1 \nu_2 \nu_3)$ as proposed by Fu and Kane[51]. The products of the parities eigenvalues at all time-reversal-invariant-momenta (TRIM) classify the bulk GaGeTe as a strong topological insulator with $\nu_0;(\nu_1 \nu_2 \nu_3) = 1;(111)$ as calculated for the primitive unit cell or, equally, with $\nu_0;(\nu_1 \nu_2 \nu_3) = 1;(001)$ as calculated for the conventional unit cell (see Table S2 in ESI for the respective parity eigenvalues at the TRIM points).

Many-body effects are known to have great impact on the gap edge states in semiconductors. Contradictorily to above, calculations with the exact exchange-correlation HSE06 functional yield a dramatically increased band gap of 550 meV in bulk GaGeTe which entails the trivial character of the electronic spectrum (calculated $Z_2=(0;000)$). It is also reflected by the changed dispersion of the valence-band edge near the T point (Fig. 2c). Furthermore, the applied GW correction also results in the transition from a topological to a trivial insulator, although the band gap expands less drastically (Table 1, Fig. S2b).

The confronting results of the calculations cannot be unequivocally juxtaposed with the published transport properties of bulk GaGeTe which rise concerns and require careful revision. Optical measurements[32] on GaGeTe crystals with GaTe impurities (documented by X-ray experiments) reported two transmittance maxima at 0.4 and 1.0 eV. The authors ascribed the first one to intra-band transitions due to *p*-doping of the sample and the second one to inter-band transitions. This finding accords in principle with the earlier mentioned band gap of 1.1 eV[9] that, nonetheless, was not supported by any experimental evidence.

In order to re-interpret the optical experimental investigation[32], we computed absorption index (k), refractive index (n), and refractivity (R) on the basis of the dielectric function obtained within the Random Phase Approximation. These quantities derived from the in-plane and out-of-plane components of the dielectric function are shown in Fig. 4a,b. Two peculiarities at ~0.2 eV and ~1 eV (denoted henceforward as A and B, respectively) are clearly visible. The former is observed for the out-of-plane component and is poorly resolved, whereas the latter is more pronounced for the in-plane component. Occurrence of these peaks can now be rationalized on the basis of the available GaGeTe band structure (see Fig. 4c). The first peak (A) corresponds to inter-band optical transitions in the area of the T point. Due to the finite size of the *q*-point mesh used in our calculations, e. g. $7 \times 7 \times 7$, the location of this peak has falsely shifted in the direction of higher energies. Based on this, the transmittance maximum registered at 0.4 eV[32] may be associated with the optical band gap, with a caveat that the earlier interpretation[32] cannot be ruled out either. The second peak (B) can be explained by inter-band optical transitions near the Γ point and is in full agreement with the previous experimental observations[32]. The discussed transitions are schematically presented in Fig. 4c. To conclude, our results offer a new interpretation of the experimental data[32] and a plausible explanation of the observed discrepancy between the bandgap estimated from the optical experiment[32] (1 eV) and from our computations (55–550 meV dependent on the chosen functional).

If GaGeTe is a trivial semiconductor as found within the HSE functional, artificial augmentation of spin-orbit strength may be considered as a means to evoke a topological phase transition from the trivial insulator into the topological insulator phase. One possible way to trigger the topological transition could thus be chemical substitution by an isovalent element with higher atomic number and, respectively, the stronger effective spin-orbit coupling interaction. In this line of thought, two hypothetical compounds, namely GaSnTe and InSnTe, were further considered. Structural optimisation within the DFT-D3 scheme has confirmed that they are isostructural to GaGeTe. Tin substitution for germanium in GaSnTe corresponds to the increment in the intrinsic spin-orbit strength parameter $\lambda_0$ from 0.29 (Ge) to 0.8 (Sn).[97] Furthermore, InSnTe possesses the largest effective spin-orbit strength in this series thanks to the $\lambda_0$ values increasing from 0.174 (Ga) to 0.392 (In).[99]

The electronic structure of GaSnTe (Fig. 3b, Fig. 5a) considered in the framework of the PBE functional demonstrates strong similarities to GaGeTe in the broad range of energies. For instance, the Dirac cone located at the W point at about −8 eV is in this case formed by the Sn *s*-orbitals and is thus attributed to the Sn bonding in the stanene-like fragment. Decisively for the present discussion, the bulk band structure of GaSnTe demonstrates qualitative differences close to the Fermi level as it retains semimetallic character when spin-orbit interaction is accounted for. Moreover, a complex inversion of four bands takes place and involves the Ga *s*-states, the Te $p_z$-orbitals, and the Sn $p_x$- and $p_y$-states. This inversion generates several local hybridization gaps between the inverted bands (Fig. 5b). As a result, emergence of surface states with topological nature can be readily expected which would enable us to classify GaSnTe as a semimetal with topological properties. Similar features were found for the Bi$_x$TeI (*x* = 2, 3) family of topological materials[100,101]. The observed metallic character is maintained with a few minor changes in the vicinity of the T point when the electronic structure is treated by the HSE exact exchange functional (Fig. 5c). Similarly to the PBE case, local hybridization gaps are also found.

As anticipated, in the case of InSnTe (Fig. 3c, Fig. 6) presence of elements with stronger effective spin-orbit coupling interaction with respect to GaGeTe leads to the topological phase transition. Both PBE and HSE exchange functionals concertedly yield an inverted energy gap in the bulk electronic structure of InSnTe. The topological $Z_2$ invariants $\nu_0;(\nu_1 \nu_2 \nu_3) = 1;(111)$ calculated from the products of the parities eigenvalues at all time-reversal-invariant-momenta (TRIM) and following the method proposed in Z2Pack[52,53] coincide with those obtained for GaGeTe in the PBE case (cf. Table S2). Nevertheless, the size and the character of the gap edge dispersion differ significantly for both parametrisations (Fig. 6b,c and Table 1). In the PBE case, an indirect bulk band gap is observed and three bands partake in the complex inversion similarly to the previously discussed GaSnTe spectrum calculated within PBE. The electronic structure of InSnTe obtained in the HSE case is characterized by a direct, small band gap of ca. 20 meV. The gap edges are also inverted but this time only two bands were involved. The analysis of the atomic composition within HSE demonstrates that InSnTe is in proximity of a transition to the trivial phase (Fig. 6c).

**Chemical bonding in *AX*Te (*A* = Ga, In; *X* = Ge, Sn) and comparison with functionalized *X*enes**

Despite notable differences in the electronic structures near the Fermi energy, chemical bonding, as evaluated by means of positional-space bonding analysis, appears very similar in all *AX*Te (the quantitative results are summarized in Tables S3–S5 of ESI) The effective charges of QTAIM atoms indicate electron transfer from gallium/indium to tellurium, while the Ge/Sn atoms remain almost neutral (Table S3, S5 in ESI). The resultant effective atomic charges for all *AX*Te generally accord with the $Ga^{2+}Ge^{0}Te^{2-}$ charge scheme derived earlier[9] from the structural considerations.

Tetrahedral atomic coordination in *AX*Te signifies strong covalent intra-layer bonding between the nearest neighbors as revealed by delocalization indices close to 1, indicating essential electron sharing between the covalently bonded atoms. For instance, $\delta(Ge,Ge) = 0.80$, $\delta(Ge,Ga) = 0.73$, $\delta(Ga,Te) = 0.70$ in GaGeTe, and $\delta(Sn,Sn) = 0.80$, $\delta(Sn,In) = 0.70$, $\delta(In,Te) = 0.67$ (cf. Table S4 in ESI).

Interactions between more distant atoms are more difficult to analyse since no ELI-D basins are present and the delocalization indices are much smaller. In the following the case of GaGeTe is considered in detail. The DI value between the next-nearest Ge atoms (2NN) from the same Ge layer are $\delta(Ge,Ge) = 0.046$ (at the distance of 4.048 Å). The DI value between two 3NN (next-next-nearest neighbors) Ge atoms is one order of magnitude smaller ($\delta(Ge,Ge) = 0.005$ for the distance of 4.736 Å). The observed ratios between the DI values resemble those in diamond ($\delta(C,C) = 0.91$ for the first coordination sphere; $\delta(C,C) = 0.039$ for the second coordination sphere; $\delta(C,C) = 0.008$ for the third coordination sphere)[82,83] as opposed to graphite (the respective values are $\delta(C,C) = 1.20$, 0.058 and 0.038)[82,83]. These findings highlight the similarity between the buckled germanium layer in GaGeTe and GeH and are consistent with the semiconducting behavior of GaGeTe.

As it follows from the charge analysis, the buckled Ge/Sn fragment within the hextuple layers of *AX*Te can be put alongside *X*anes[7] as yet another example of *X*ene functionalization. Notably, geometrical characteristics of the germanium sheet (interatomic distances, bonding angles and buckling) are almost identical in GaGeTe and GeH, and bear salient structural similarities to the respective fragment in the Zintl compound □-CaGe$_2$ which acted as the precursor for germanane[28] (Fig. 7, Table 2). Similar tendencies are observed in the series InSnTe (GaSnTe) – SnH (stanane)[102] – BaSn$_2$ (Table 2), although it should be noted that stanane has not been experimentally characterized yet. On the other hand, the chemical bonding scenarios differ in these series of compounds; we again depict it on the example of germanium. The negatively charged germanium sheet with slightly longer Ge–Ge distances and more pronounced buckling than in germanene (cf. Table 2) is incorporated in an ionic arrangement of Ca$^{2+}$ cations in *β*-CaGe$_2$, whereas its analogue is covalently functionalized in GeH and GaGeTe. Two types of functionalization entail prominently different electronic properties. Ionic functionalization in *β*-CaGe$_2$ results in a semimetallic ground state, whereas germanane which is covalently functionalized via hydrogenation is a topologically trivial wide-gap semiconductor,[16] and its band-gap size can be flexibly varied by chemical pressure up to 15 %.[30] Similarly, stanane[103] is expected to be trivial, whereas halogen-functionalized stanene[102] is theoretically predicted to be topological. In GaGeTe, which exhibits covalently bonded, almost neutral germanium sheets sandwiched between the GaTe fragments, a non-conducting ground state is realized. Given stronger spin-orbit coupling interaction, like in InSnTe, transition into the topological state may occur.

Unlike the considered GaGeTe-type compounds with formally neutral hextuple layers, topological Zintl phases $MX_2$[23–25] (*M* = Ca, Sr, Ba; *X* = Si, Ge, Sn) are bound to uncompensated surface charge upon cleaving, which interferes with the observations of the topological states by e. g. transport measurements. These hindering effects were examined in details, for instance, for a weak topological insulator built by alternating charged layers.[104] Hence the GaGeTe-type topological materials may be much easier to handle than Zintl phases.

**CONCLUSIONS**

First-principles calculations identify the GaGeTe-type periodic structures as a potential host for topological phases. The layered InSnTe bulk material is predicted to be a 3D strong topological insulator with $Z_2 = 1;(111)$. Unlike structurally related *X*enes (2D TI) or stanene-based $BaSn_2$ (3D TI), which valence and conduction bands are dominated by atomic-orbital contributions of the buckled honeycomb structural fragment, the complex band inversion in InSnTe, as found within the PBE functional, is realized by the In-$5s$ states, Sn-$5s$ and Sn-$5p$ states of the *X*ene-like sheets and the SOC-splitted Te-$5p$ states. Thus, covalent functionalization of the *X*ene-like building block in the periodic 3D stack of the GaGeTe-type implicates a topological state. Experimental confirmation is urgently called for hypothetical GaSnTe and InSnTe materials which are shown to be dynamically stable.

As far as the series forerunner GaGeTe is concerned, transport experiments and spectroscopy studies are currently underway to confront the contradicting theoretical predictions. The measurements on GaGeTe appear feasible thanks to its high stability in contrast to air-sensitive Zintl compounds and artificial 2D materials. In contrast to germanane which quickly becomes amorphous above 75 °C,[28] GaGeTe offers both thermal (melts peritectically at 800 °C[9]) and chemical (resistant to air, water and NaOH(aq)[9]) stability.

Although the tetrahedral atomic coordination in the GaGeTe-type structures closely resembles the topological materials with the diamond-like cubic lattices[105–107], there is no direct similarity between the inversion mechanisms in these two groups. Furthermore the GaGeTe structure cannot be derived directly from a diamond-like 3D lattice. Unlike HgTe-based topological insulators with isotropic diamond-like cubic lattices, GaGeTe-type features van der Waals gaps and is thus a promising candidate for engineering of superlattices, innately related to PbTe, SnTe, HgTe, GeTe, GaAs, etc. Being composed of accessible elements, which are neither refractory nor too volatile or corrosive, GaGeTe-type may be suitable for thin film manufacture.

Chemical modification of the parent GaGeTe compound seems feasible. One of the possible ways to induce topological order could be partial doping with larger isovalent *p*-elements, and structurally-related zincblende-type semiconductors offer a vast playground for that. Furthermore, effects of magnetic doping as well as intercalation of magnetic dopants into van der Waals gaps on the topological properties can be explored further.

**Author Contributions**

The manuscript was written by A.I., F.P., T.V.M. and I.P.R. through the contributions of all authors. F.P., T.V.M., I.P.R., A.I.B. performed DFT and GW calculations and analyzed the data under supervision of A.P. and E.V.C. F.P., A.I., A.Z. and M.R. analyzed the crystallochemical data. I.Y.S. and I.P.R. calculated and analyzed the optical properties and phonon spectra under supervision of R.H., K.-P.B. and E.V.C. P.G., A.I.B. and A.I. performed chemical bonding analysis. F.P., T.V.M. and A.I. supervised and coordinated the project.

**Acknowledgements**

F. P., A. I., A. Z., A. P., and M. R. acknowledge the financial support by the German Research Foundation (DFG) in the framework of Special Priority Programme "Topological insulators" (SPP 1666) and ERANET-Chemistry. T. V. M. and I. P. R. acknowledge support by "Tomsk State University competitiveness improvement programme" and the Saint Petersburg State University project No. 15.61.202.2015. I. P. R. acknowledges support by the Ministry of Education and Science of the Russian Federation within the framework of the governmental program "Megagrants" (state task No. 3.8895.2017/П220). T. V. M. and I. P. R.

have performed the calculations on the SKIF-Cyberia supercomputer of Tomsk State University. E. V. C. acknowledges the Spanish Ministry of Science and Innovation (grants no. FIS2013-48286-C02-02-P and FIS2013-48286-C02-01-P) and the Basque Departamento de Educacion, UPV/EHU (grant IT-756-13). P. G. and A. B. acknowledge the German Research Foundation (grant BA-4911/1-1) and ZIH TU Dresden for the provided computational facilities.

The authors are indebted to R. Weihrich (University of Augsburg, Institute of Materials Resource Management) and V. M. Silkin (Donostia International Physics Center (DIPC)) for fruitful discussions and critical reading.

**FIGURES AND TABLES**

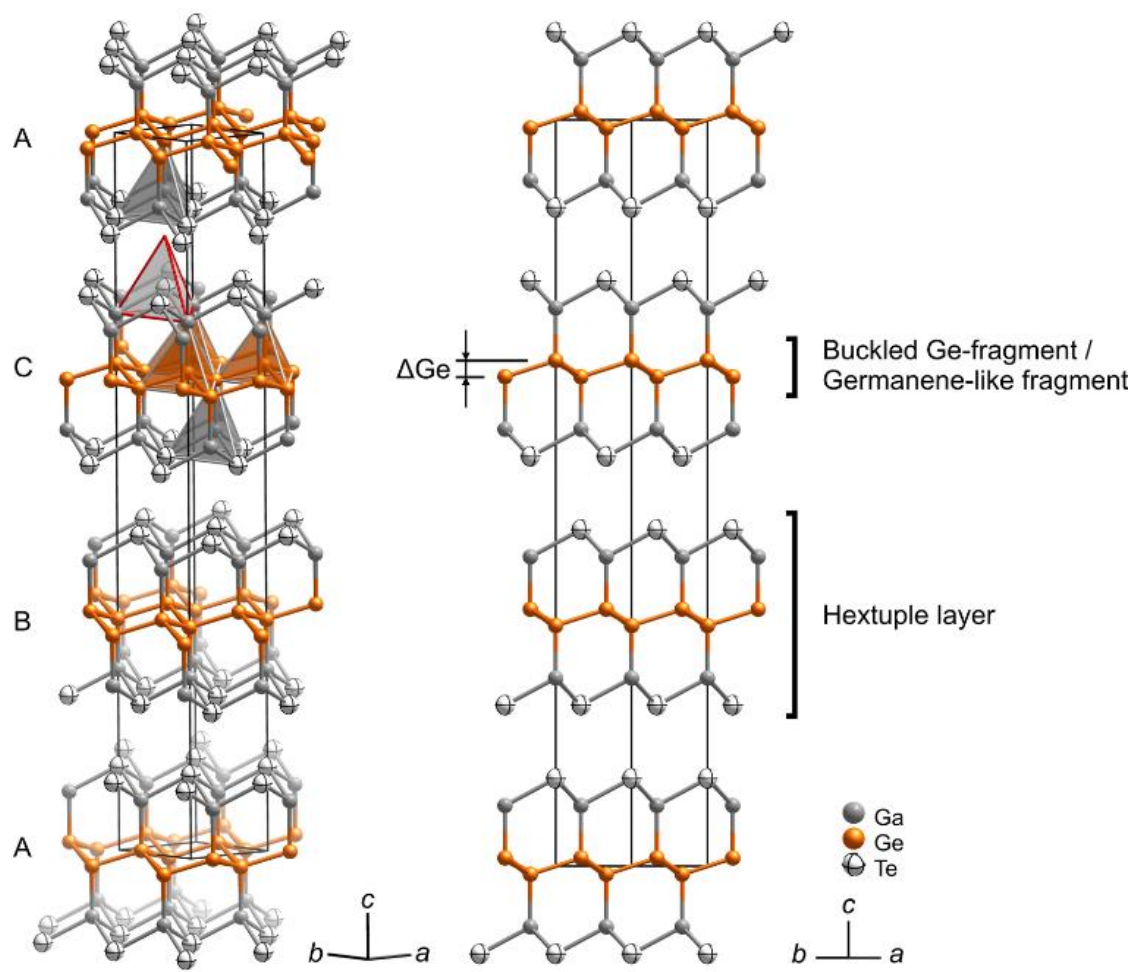

**Figure 1.** Selected views of the bulk GaGeTe structure (conventional unit cell outlined) with the notions used for the structural fragments in the text. The coordination polyhedra emphasize structural relations with a diamond-like lattice. Note the missing vertex of the Te-centered polyhedron (see text). ΔGe defines the buckling of the Ge fragment as a normal between two Ge atomic planes.

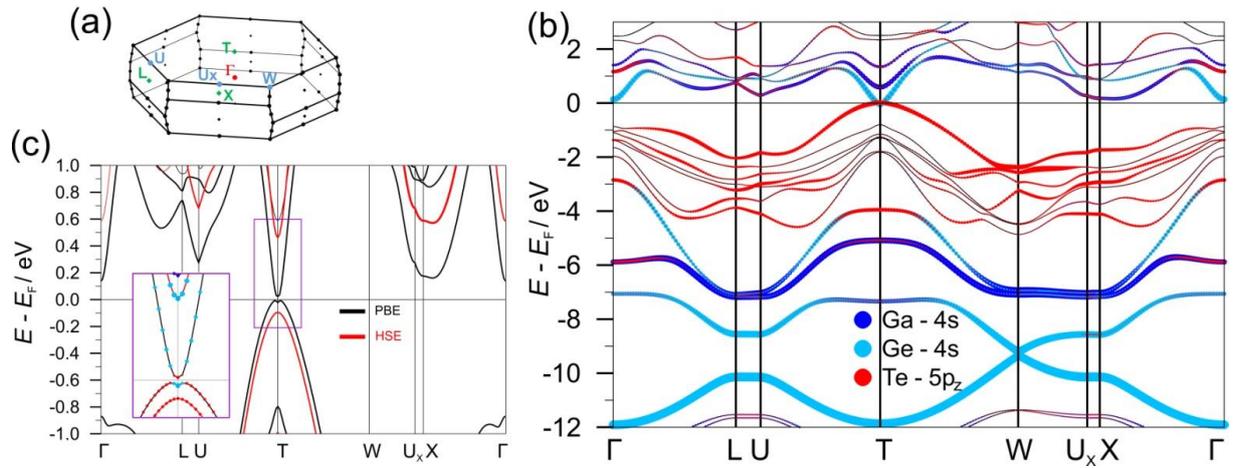

**Figure 2.** (a) 3D Brillouin zone for the primitive unit cell of GaGeTe. (b) Bulk band structure of GaGeTe with spin-orbit coupling. In the panel (c) the bulk band structure calculated within PBE and HSE functional is blown up in the vicinity of the Fermi level. The color-coding of the atomic contributions is identical in both panels. Filled circles denote atomic compositions with the *s*- and $p_z$-symmetry for the Ga, Ge and Te atoms, respectively.

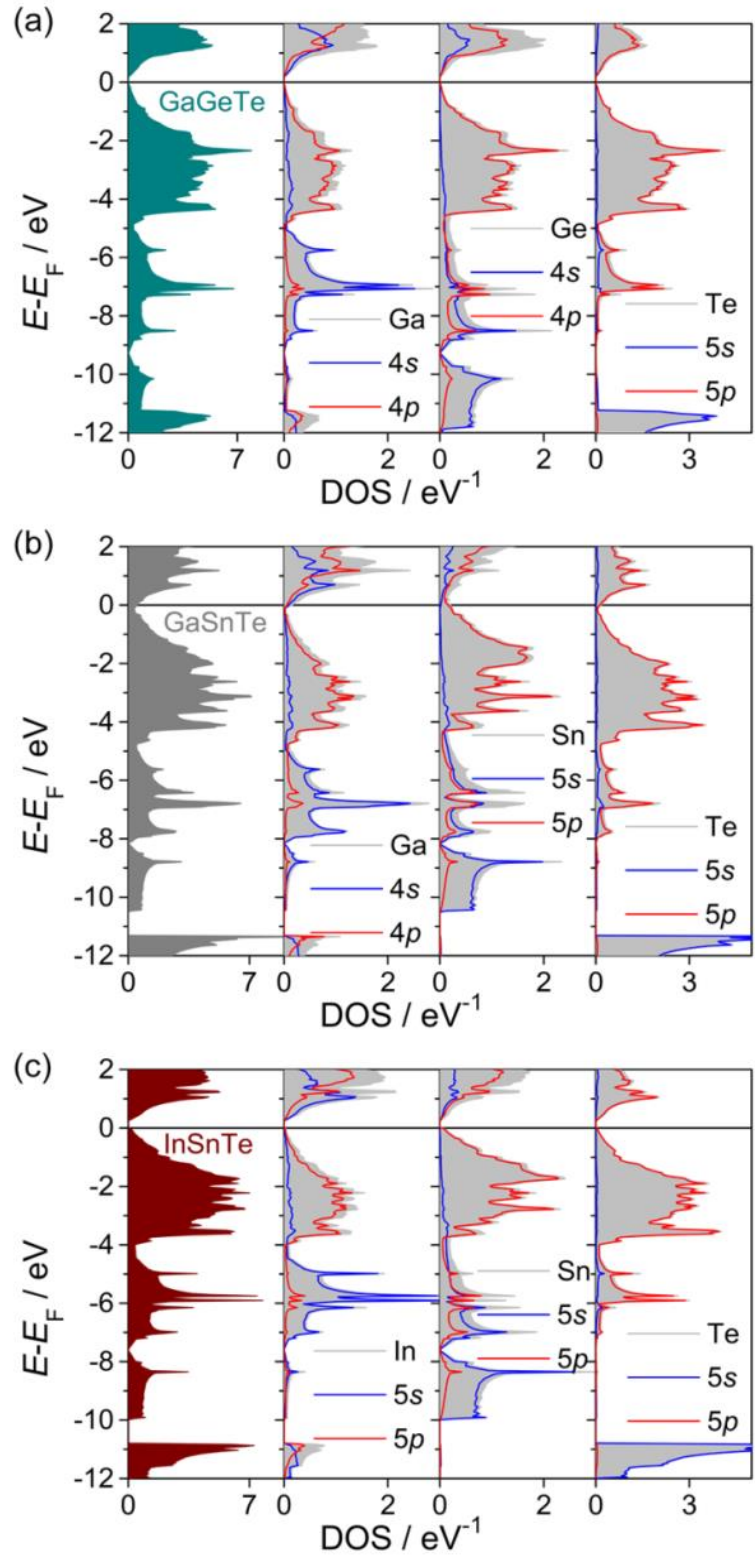

**Figure 3.** DOS plots for GaGeTe (a), GaSnTe (b) and InSnTe (c) with the atomic orbital projected-DOS (full-relativistic FPLO-LDA).

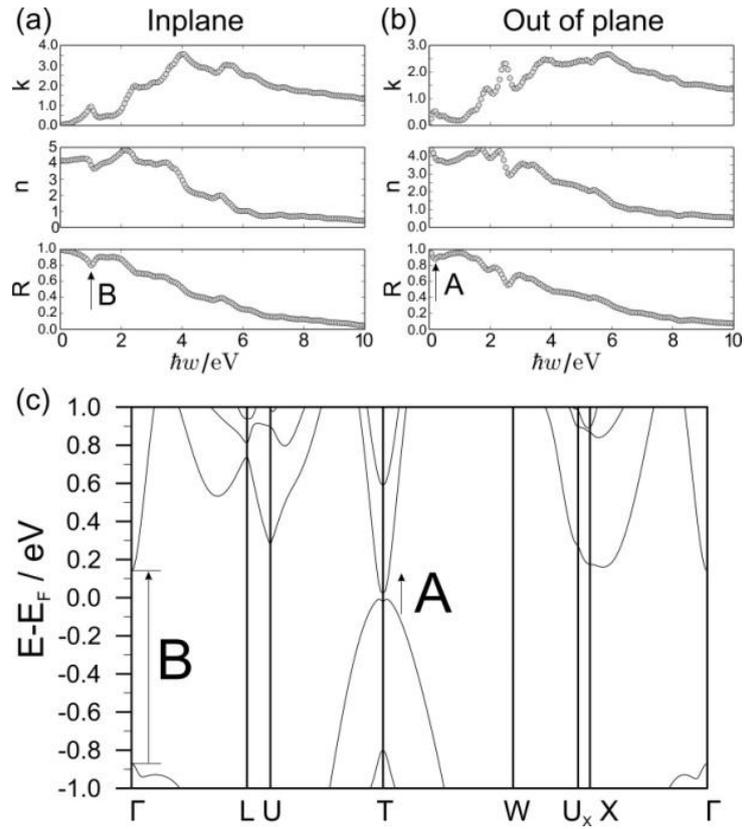

**Figure 4.** Absorption index (k), refractive index (n), and refractivity (R) calculated for GaGeTe on the basis of the in-plane (a) and out-of-plane (b) components of the dielectric function. (c) Bulk electronic structure of GaGeTe near the Fermi level. The arrows define the optical transitions corresponding to the peaks in (a) and (b).

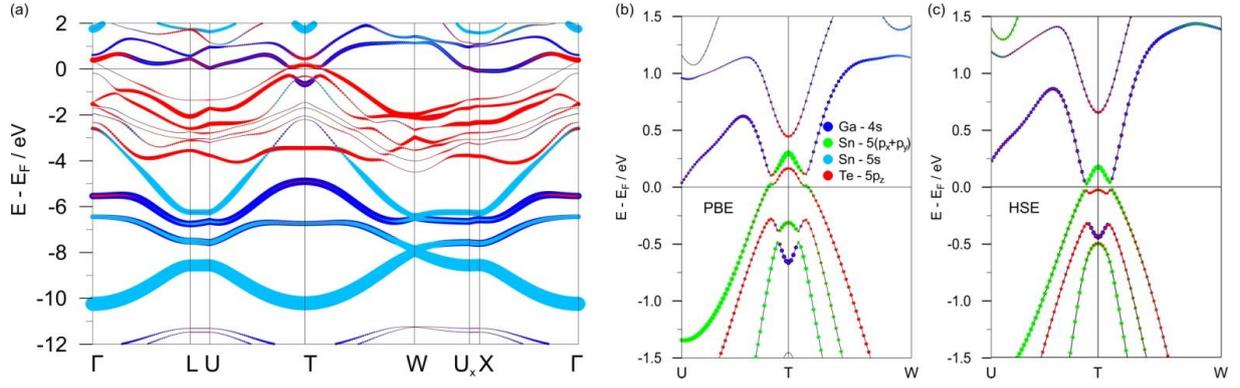

**Figure 5.** Bulk band structure of GaSnTe with spin-orbit coupling. The color-coding for the atomic contributions is identical in both panels. Filled circles correspond to the atomic compositions with $s$-, $p_{x+y}$- and $p_z$- symmetry for Ga, Sn and Te.

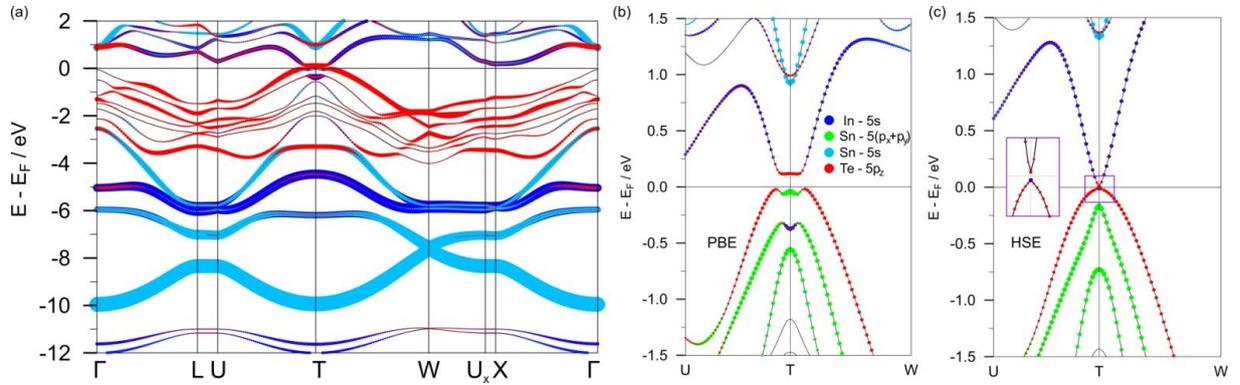

**Figure 6.** Bulk band structure of InSnTe with spin-orbit coupling. The color-coding for the atomic contributions is identical in both panels. Filled circles correspond to the atomic compositions with $s$-, $p_{x+y}$- and $p_z$- symmetry for In, Sn and Te.

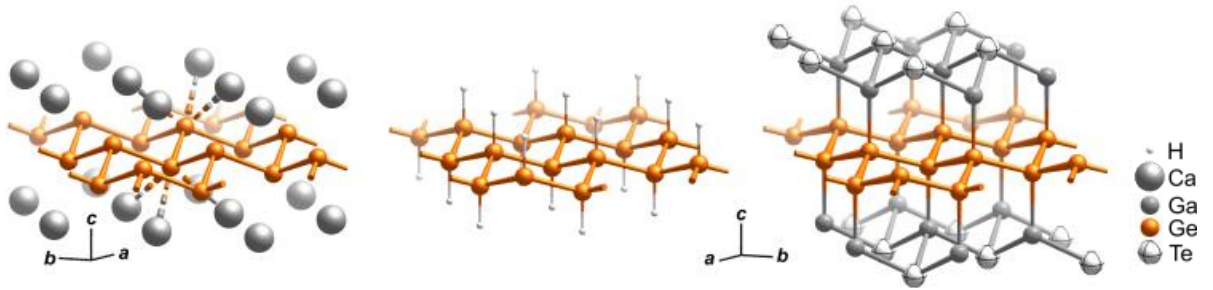

**Figure 7.** Layered fragments of crystal structures of $\beta$-CaGe$_2$,[108] (left) and GaGeTe (right) compared to germanane GeH (center). Atomic coordinates of GeH are taken from[28,29]). The *X*ene (i. e. germanene in this case) fragment is highlighted by yellow colour. Its geometrical characteristics are summarized in Table 2 together with the respective data on stanene.

**Table 1.** A summary of selected optimized geometrical parameters and electronic band gaps (SOC included) calculated for the bulk *AX*Te (*A* = Ga, In; *X* = Ge, Sn) using various DFT-based codes and parametrisation. Relaxation was performed under space-group restrained (no. 166), whereas the unit cell parameters and atomic coordinates were allowed to vary. Since structure relaxation cannot be performed with the HSE06 functional in VASP, the geometry obtained with the PBE functional was used instead.

| Computational details (method–functional) | $a$ / Å | $c$ / Å | $V$ / Å$^3$ | inter-layer $d$ / Å | $d(X–X)$ / Å | $d(X–A)$ / Å | $d(A–Te)$ / Å | ∠ $X–X–X$ / ° | $\Delta X$ / Å | Band gap / meV |
|---|---|---|---|---|---|---|---|---|---|---|
| Experimental geometry of GaGeTe[7] | | | | | | | | | | |
| FPLO–LDA* | 4.048 | 34.731 | 492.87 | 3.408 | 2.457 | 2.442 | 2.657 | 110.90 | 0.759 | 57 |
| FPLO–PBE* | | | | | | | | | | 29 |
| FPLAPW–LDA** | | | | | | | | | | 70 |
| PAW–PBE*** | | | | | | | | | | 33 |
| Optimized geometry of GaGeTe (cf. Table S1) | | | | | | | | | | |
| FPLO–LDA* | 4.027 | 34.400 | 483.16 | 3.295 | 2.458 | 2.415 | 2.650 | 110.02 | 0.797 | 21 |
| PAW–PBE+D3*** | 4.086 | 34.583 | 500.11 | 3.271 | 2.491 | 2.451 | 2.683 | 110.20 | 0.800 | 55 |
| PAW–HSE*** | 4.086 | 34.583 | 500.11 | 3.271 | 2.491 | 2.451 | 2.683 | 110.20 | 0.800 | 550 |
| GW | 4.086 | 34.583 | 500.11 | 3.271 | 2.491 | 2.451 | 2.683 | 110.20 | 0.800 | 298 |
| Optimized geometry of GaSnTe (cf. Table S1) | | | | | | | | | | |
| PAW–PBE+D3*** | 4.318 | 35.754 | 577.33 | 3.172 | 2.746 | 2.631 | 2.752 | 103.68 | 1.151 | 0 |
| PAW–HSE*** | 4.318 | 35.754 | 577.33 | 3.172 | 2.746 | 2.631 | 2.752 | 103.68 | 1.151 | 0 |
| LCAO-LDA/FPLO-LDA**** | 4.283 | 34.984 | 555.76 | 3.112 | 2.706 | 2.578 | 2.726 | 104.65 | 1.098 | 0 |
| LCAO-PBE+D2/FPLO-PBE**** | 4.296 | 35.372 | 565.48 | 3.123 | 2.730 | 2.592 | 2.743 | 103.79 | 1.140 | 0 |
| Optimized geometry of InSnTe (cf. Table S1) | | | | | | | | | | |
| PAW-PBE+D3*** | 4.482 | 37.113 | 645.64 | 3.111 | 2.784 | 2.805 | 2.901 | 107.22 | 1.026 | 137 (indirect) |
| PAW-HSE*** | 4.482 | 37.113 | 645.64 | 3.111 | 2.784 | 2.805 | 2.901 | 107.22 | 1.026 | 20 (direct) |
| LCAO-LDA/FPLO-LDA**** | 4.450 | 36.766 | 630.38 | 3.069 | 2.775 | 2.788 | 2.871 | 106.61 | 1.049 | 8 |
| LCAO-PBE+D2/FPLO-PBE**** | 4.465 | 37.321 | 644.35 | 3.058 | 2.796 | 2.819 | 2.901 | 105.99 | 1.082 | 150 (indirect) |

\* FPLO software package,[67] \*\* ELK software package,[69] \*\*\* VASP software package[41–43]. \*\*\*\* structure optimization with CRYSTAL[56], band structure calculated with FPLO. $\Delta X$ defines the height of the buckled germanium / tin fragment (cf. Fig. 1).

**Table 2.** Comparison of geometric characteristics of germanene/stanene–like structural fragments incorporated in selected 2D *X*anes, Zintl phases and *AX*Te materials. Degree of buckling Δ*X* is defined as a normal between two *X* atomic planes.

| Material | d(X–X) / Å | ∠ X–X–X / ° | ΔX / Å |
|---|---|---|---|
| Germanene-like structural fragment | | | |
| β-CaGe$_2$[108] | 2.519 | 104.6 | 1.02 |
| GeH[28,29] | 2.435 | 109.8 | 0.80 |
| GaGeTe[9] | 2.457 | 110.9 | 0.76 |
| Stanene-like structural fragment | | | |
| BaSn$_2$ | 2.919 | 105.66 | 1.14 |
| SnH[109] | 2.88 | – | 1.2 |
| GaSnTe (this work) | 2.746 | 103.68 | 1.15 |
| InSnTe (this work) | 2.784 | 107.22 | 1.03 |